\documentclass[12pt]{article}
\usepackage[utf8]{inputenc}
\usepackage{amsmath,ulem} 
\usepackage[mathscr]{eucal}
\usepackage{color}
\input{epsf} 
\newcommand\as{\alpha_{\mathrm{S}}} 
\newcommand\f[2]{\frac{#1}{#2}}

\def\beq{\begin{equation}} 
\def\eeq{\end{equation}} 
\def\beeq{\begin{eqnarray}} 
\def\eeeq{\end{eqnarray}} 
 
\def\to{\rightarrow} 
\def\nn{\nonumber}

\def\ms{${\overline {\rm MS}}$}

\def\b0{\beta_0}

\def\GE{\gamma_E}
\def\ep{\epsilon}

\def\At{{\widetilde {\mathscr A}}_{i}}

\def\A{{{\mathscr A}}_{i}}

\def\ATt{{\widetilde {\mathscr A}}_{T,i}}

\def\AT{{{\mathscr A}}_{T,i}}

\def\Aot{{\widetilde {\mathscr A}}_{0,i}}
  
\def\Ao{{{\mathscr A}}_{0,i}}

\def\Anoti{{{\widetilde{\mathscr A}}}}

\def\Anott{{{\mathscr A}}}

\begin{document} 
\begin{titlepage}
\begin{flushright}
ZU-TH 17/19\\
ICAS 42/19\\
\end{flushright}

\renewcommand{\thefootnote}{\fnsymbol{footnote}}
\vspace*{0.5cm}

\begin{center}
{\large \bf Soft-gluon effective coupling \\[0.3cm] and cusp anomalous dimension}
\end{center}

\par \vspace{2mm}
\begin{center}
{\bf Stefano Catani${}^{(a)}$,
Daniel de Florian${}^{(b)}$
and
Massimiliano Grazzini${}^{(c)}$}

\vspace{5mm}

${}^{(a)}$INFN, Sezione di Firenze and
Dipartimento di Fisica e Astronomia,\\ 
Universit\`a di Firenze,
I-50019 Sesto Fiorentino, Florence, Italy

${}^{(b)}$
International Center for Advanced Studies (ICAS), ECyT-UNSAM,\\
Campus Miguelete, 25 de Mayo y Francia, (1650) Buenos Aires, Argentina

${}^{(c)}$Physik Institut, Universit\"at Z\"urich, CH-8057 Z\"urich, Switzerland

\vspace{5mm}

\end{center}

\par \vspace{2mm}
\begin{center} {\large \bf Abstract} 

\end{center}
\begin{quote}
\pretolerance 10000

We consider
the extension of the
CMW soft-gluon effective coupling \cite{Catani:1990rr}
in the context of soft-gluon resummation for QCD hard-scattering observables
beyond the next-to-leading logarithmic accuracy.
We present two proposals of a soft-gluon effective coupling that extend the CMW coupling to all perturbative orders in the \ms\ coupling $\as$. 
Although both effective couplings are well-defined in the physical four-dimensional space time, we examine their behaviour
in $d=4 -2\ep$ space time dimensions. 
We uncover an all-order perturbative relation with
the cusp anomalous dimension: the (four dimensional) cusp anomalous dimension is equal to the $d$-dimensional soft-gluon effective coupling
at the conformal point $\ep=\beta(\as)$, where the $d$-dimensional QCD $\beta$-function, $\beta(\as) - \ep$, vanishes.
We present the explicit expressions of the two soft-gluon couplings
up to ${\cal O}(\as^2)$ in $d$ dimensions. 
In the four-dimensional case we compute the two soft couplings up to 
${\cal O}(\as^3)$.
For one of the two couplings, we confirm the ${\cal O}(\as^3)$ result previously presented by other authors.
For the other coupling, we obtain the explicit relation with the cusp anomalous dimension up to ${\cal O}(\as^4)$.
We comment on Casimir scaling at ${\cal O}(\as^4)$.

\end{quote}

\vspace*{\fill}
\begin{flushleft}
April 2019
\end{flushleft}
\end{titlepage}

\setcounter{footnote}{1}
\renewcommand{\thefootnote}{\fnsymbol{footnote}}

A well known feature of QCD is that perturbative computations of hard-scattering processes are sensitive to soft-gluon effects.
These effects manifest themselves 
in hard-scattering observables that are
evaluated close to the exclusive boundary of the phase space.
In such kinematical configurations, real-radiation contributions in the inclusive final state are strongly suppressed and they cannot balance virtual-radiation effects (which are always kinematically allowed).
The unbalance leads to large logarithmic radiative corrections (the argument of the logarithms is the distance from the exclusive boundary).
These contributions are often refereed to as logarithmically-enhanced terms of Sudakov type.

Real emission has a logarithmic spectrum for radiation that is {\it soft} and/or
{\it collinear} to the direction of the observed hard jets (partons). This spectrum produces perturbative contributions that have at most two powers of logarithms
for each power of the QCD coupling $\as$. In the case of many observables, the 
double-logarithmic (DL) terms can be resummed to all orders in $\as$ in 
{\it exponentiated} form. For most of these observables resummation can be extended in exponentiated form 
to higher (or arbitrary) logarithmic orders. This feature depends on the hard-scattering process and, especially, on the specific kinematical properties of the observable under consideration. In the case of observables that fulfil exponentiation, it is natural to device a resummed perturbative expansion by systematically organizing the exponent in classes of subsequent logarithmic accuracy: 
leading logarithmic (LL) terms, next-to-leading logarithmic (NLL) terms,
next-to-next-to-leading logarithmic (NNLL) terms and so forth.

The explicit computation and resummation of Sudakov-type logarithms can be performed by using both traditional perturbative QCD methods and techniques based on Soft Collinear Effective Theory. Very many observables are nowadays known up to NLL accuracy (see, e.g., the reviews in Refs.~\cite{Luisoni:2015xha, Becher:2014oda}
and references therein for a large, though still incomplete, list of NLL results),
and several observables are known to NNLL or higher logarithmic accuracy
(a list of results can be found in Ref.~\cite{Banfi:2018mcq}).

Direct inspection of NLL resummed results shows that they have a high degree of universality, with general structures and ingredients that have a `minimal' dependence
on the hard-scattering process and on the specific observable to be treated.
Roughly speaking, the resummed logarithmic contributions are embodied in a `generalized' Sudakov form factor and they are combined with `hard' (non-logarithmic) factors, which are computable at fixed perturbative orders.
The universality structure of NLL resummation is evident in the context of 
process-independent and observable-independent formulations of resummation that have been explicitly worked out \cite{Banfi:2003je, Bonciani:2003nt, Banfi:2004yd}
for large classes of (properly specified) observables. In particular, within such formulations, the Sudakov form factor is obtained by integration (over an observable-dependent phase space) of a universal kernel that is explicitly evaluated up to NLL accuracy. In view of these NNL results, progress is being carried out
\cite{Banfi:2014sua} to extend such observable-independent resummation program to NNLL accuracy. Obviously, an improved understanding of NNLL contributions and their possible universality features is also relevant in the context of resummed calculations for specific observables, independently of any observable-independent treatment.

A relevant feature of the NLL results that we have briefly recalled is that the `dominant' (soft {\it and} collinear) part of the NLL kernel of the generalized Sudakov form factor is obtained simply and in complete form through the use of the QCD coupling $\as^{CMW}$ \cite{Catani:1990rr}
in the Catani--Marchesini--Webber (CMW) scheme
(or bremsstrahlung scheme). The CMW coupling $\as^{CMW}$ has the meaning of an effective (physical) coupling for {\it inclusive} radiation of soft {\it and} collinear gluons. The purpose of the present paper is to extend the definition 
of $\as^{CMW}$ beyond NLL accuracy. A definition of such extension has been proposed in Ref.~\cite{Banfi:2018mcq}. Moreover, the authors of Ref.~\cite{Banfi:2018mcq}
present the relation between the effective coupling and the customary \ms\
renormalized coupling up to ${\cal O}(\as^3)$, and they embody the effective coupling in the context of an explicit formulation of NNLL resummation for generic two-jet
observables in $e^+e^-$ annihilation \cite{Banfi:2014sua}.
We think that there is no unique extension of $\as^{CMW}$ beyond NLL accuracy.
By unique extension, we mean an extension with the same universality features as those of $\as^{CMW}$ at NLL accuracy (we postpone additional comments on this).
Therefore, in the present paper, besides considering the definition of 
Ref.~\cite{Banfi:2018mcq}, we propose a different definition of the soft-gluon effective coupling, and we present some results for both definitions.

We recall that the CMW coupling plays a role in two other contexts directly or indirectly related to Sudakov resummation. The coupling $\as^{CMW}$ can be used
in Monte Carlo event generators (see, e.g., Ref.~\cite{Buckley:2011ms})
to improve the logarithmic accuracy of corresponding parton shower algorithms
\cite{Catani:1990rr}. The dispersive approach to power-behaved terms in QCD hard processes \cite{Dokshitzer:1995qm, Dokshitzer:1998pt} uses $\as^{CMW}$ to combine
contributions from the low-momentum (non-perturbative) region with perturbative
contributions at next-to-leading order (NLO) in the high-momentum region.
The extension of $\as^{CMW}$ beyond NLL accuracy can be useful also for higher-order
studies within these two contexts.

The outline of the paper is as follows. We first recall (in a sketchy way) the role 
of the CMW coupling in NLL resummed calculations. Then we introduce two definitions of the soft-gluon effective coupling at arbitrary perturbative orders. We illustrate
various perturbative results for both effective couplings, 
and we present some brief comments on their derivation (details about the derivation of the results will appear in a separate publication).
The results include 
an all-order relation with the cusp anomalous dimension
and perturbative expressions up to ${\cal O}(\as^4)$.
Finally, we conclude with a summary and some general comments.

At the lowest perturbative order, the probability of radiation of a single soft gluon
that is soft ($\omega \ll E$) and collinear ($\theta \ll 1$) to the direction of a
massless hard parton is given by the well-known DL spectrum
\begin{equation}
\label{wdl}
  dw_i^{DL}=C_i \f{\as}{\pi} \;\f{d\omega}{\omega} \;\f{d\theta^2}{\theta^2}
\simeq C_i \f{\as}{\pi} \;\f{dz}{1-z} \;\f{dq_T^2}{q_T^2} \;\;,
\end{equation}
where $\omega$ is the energy of the soft gluon, $E$ is the energy of the radiating
hard parton and $\theta$ is the gluon emission angle. To DL accuracy, $dw_i^{DL}$
can be equivalently expressed in terms of the longitudinal-momentum fraction
$1-z$ ($1-z \simeq \omega/E$) and transverse momentum $q_T$ 
($q_T \simeq \omega \theta$) of the soft gluon. The subscript $i$ denotes the type of radiating parton ($i=q,{\bar q}, g$), and $C_i$ is the corresponding quadratic Casimir coefficient. 
We have $C_i=C_F$ if $i=q$ (or $i={\bar q})$) and $C_i=C_A$ if $i=g$, with $C_F=(N_c^2-1)/(2N_c)$ and $C_A=N_c$ in $SU(N_c)$ QCD with $N_c$ colours.

The integration of the spectrum in Eq.~(\ref{wdl}) over the observable-dependent phase space produces large DL terms (in the vicinity of the exclusive boundary)
and infrared divergent contributions that are cancelled by one-loop virtual-radiation effects. In the case of Sudakov sensitive observables that fulfil exponentiation,
DL resummation is achieved by simply using $dw_i^{DL}$ as integration kernel in the exponent of the observable-dependent Sudakov form factor.

The intensity of soft-gluon radiation in Eq.~(\ref{wdl}) is $C_i \as/\pi$.
The NLL resummation of the contributions from soft {\it and} collinear radiation is obtained (see, for instance, Eqs.~(10), (12) and (26) in Ref.~\cite{Bonciani:2003nt},
or Eqs.~(2.16) and (2.29) in Ref.~\cite{Banfi:2004yd}) by using the DL kernel of 
Eq.~(\ref{wdl}) and simply replacing the intensity of the soft-gluon coupling as follows
\beq
\label{cmw}
C_i \,\f{\as}{\pi} \to \A^{CMW}(\as(q_T^2))= C_i \,\f{\as^{CMW}(q_T^2)}{\pi}
= C_i \,\f{\as(q_T^2)}{\pi}\left(1+\f{\as(q_T^2)}{2\pi} K\right)\;,
\eeq
where $\as^{CMW}$ is the CMW coupling \cite{Catani:1990rr} and $\as(\mu^2)$ is the QCD running coupling at the renormalization scale $\mu$ in the \ms\ renormalization scheme. The value of the coefficient $K$ in Eq.~(\ref{cmw}) is ($n_F$ is the number of massless-quark flavours)
\beq
\label{kcoef}
K=\left(\f{67}{18}-\f{\pi^2}{6}\right) C_A -\f{5}{9}n_F \;\;,
\eeq
as it turned out since early works on NLL resummation of several observables
\cite{Kodaira:1981nh, Catani:1988vd, Catani:1989ne}.

Two effects are embodied in the DL kernel $dw_i^{DL}$ through the 
replacement in Eq.~(\ref{cmw}).
The QCD coupling $\as$ is evaluated at the scale of the soft-gluon transverse momentum
$q_T$ \cite{Amati:1980ch, Dokshitzer:1978hw}: this accounts for the resummation of the LL terms. The gluon coupling acquires a correction of ${\cal O}(\as^2)$
(which is controlled by the coefficient $K$ in the \ms\ renormalization scheme): this produces the resummation of NLL terms.
Since the replacement takes place in the exponent of the Sudakov form factor,
it is produced by the {\it correlated} radiation of soft partons
(both two soft gluons and a soft $q{\bar q}$ pair), whereas the independent emission 
of soft gluons is taken into account through the exponentiation.
We also note that $\A^{CMW}(\as)$ is an effective coupling at the inclusive level,
since it is obtained by integrating over the momenta of the final-state correlated partons. The coupling $\A^{CMW}(\as)$ refers to radiation that is both soft and collinear. The Sudakov form factor includes other NLL terms due to soft wide-angle (i.e., non-collinear) radiation and hard (i.e., non-soft) collinear radiation: we postpone some comments on these terms.

Since the CMW Sudakov kernel refers to soft and collinear radiation, it can be viewed
as obtained by considering the soft limit of multiple collinear radiation.
In this respect it is natural to compare it with
the DGLAP kernel \cite{Moch:2004pa} that controls the collinear evolution of the parton distribution functions (PDFs). In the soft limit, $z \to 1$, the flavour diagonal DGLAP kernel $P_{ii}(\as;z)$ ($1-z$ is the longitudinal-momentum fraction
that is radiated in the final state) has the following behaviour
\cite{Korchemsky:1988si}:
\beq
\label{ap}
P_{ii}(\as;z) = \frac{1}{1-z} \;A_i(\as) + \dots \;\;,  \quad \quad \quad (z < 1) \;,
\eeq
where the dots on the right-hand side denote terms that are less singular than 
$(1-z)^{-1}$ (we have also neglected contact terms, proportional to $\delta(1-z)$,
of virtual origin).
The soft behaviour in Eq.~(\ref{ap}) also applies to the collinear evolution of the parton fragmentation functions. 

The perturbative function $A_i(\as)$ in 
Eq.~(\ref{ap}) is usually called (light-like) cusp anomalous dimension, since it can also be related to the renormalization of cusp singularities of Wilson loops 
\cite{Brandt:1981kf, Korchemsky:1987wg}. In the context of our discussion,
$A_i(\as)$ directly refers to the soft limit in Eq.~(\ref{ap}), independently of any relations with Wilson loop renormalization. The perturbative expansion of $A_i(\as)$
reads
\begin{equation}
\label{cusp}
 A_i(\as)=\sum_{n=1}^\infty \left(\frac{\as}{\pi}\right)^n A_i^{(n)}\, . 
\end{equation}
where $\as$ is the renormalized \ms\ coupling. 
The perturbative coefficients $A^{(1)}_i, A^{(2)}_i$ 
\cite{Curci:1980uw, Furmanski:1980cm} and $A_i^{(3)}$ \cite{Moch:2004pa} are explicitly known. 
Using the \ms\ {\it factorization} scheme for PDFs and fragmentation functions, these coefficients are
\begin{align}
\label{a1cusp}
A_i^{(1)}&=C_i \;\;,\\
\label{a2cusp}
A_i^{(2)}&=\f{1}{2}K\,C_i \;\;,\\
A_i^{(3)}&=C_i
\Bigg[
\left(\f{245}{96}-\f{67}{216}\pi^2+\f{11}{720}\pi^4+\f{11}{24}\zeta_3\right)C_A^2+\left(-\f{209}{432}+\f{5}{108}\pi^2-\f{7}{12}\zeta_3\right)\
C_A\, n_F\nn\\
&+\left(-\f{55}{96}+\f{1}{2}\zeta_3\right)C_F\, n_F-\f{1}{108}n_F^2
\; \Bigg]
\;,
\label{a3cusp}
\end{align}
where $\zeta_k$ is the Riemann $\zeta$-function.
The fourth-order coefficient $A^{(4)}_i$ is known in approximate numerical form
\cite{Moch:2017uml, Moch:2018wjh}
(the calculation in full analytic form is under completion),
and a first numerical 
estimate of $A_q^{(5)}$ has been presented recently \cite{Herzog:2018kwj}.
By direct inspection of Eqs.~(\ref{a1cusp})--(\ref{a3cusp})
we note that the dependence on $i$ 
(the type of radiating parton)
of the perturbative function $A_i(\as)$
is entirely specified up to ${\cal O}(\as^3)$
by the overall colour factor $C_i$. This overall dependence on $C_i$,
which is customarily named as {\it Casimir scaling} relation,
follows from the soft-parton origin of $A_i(\as)$ \cite{Gatheral:1983cz}, and it is violated 
at higher perturbative orders \cite{Frenkel:1984pz}, starting from 
${\cal O}(\as^4)$.

From Eqs.~(\ref{cusp})--(\ref{a2cusp}) we see that,
up to the second perturbative order, $A_i(\as)$ coincides with the CMW coupling
$\A^{CMW}(\as)$ in Eqs.~(\ref{cmw}) and (\ref{kcoef}).
One may be tempted to conclude that
the cusp anomalous dimension provides a sensible definition of a physical
(though effective) soft-gluon coupling beyond ${\cal O}(\as^2)$. 
The equivalence between $A_i(\as)$ and soft-gluon coupling, however, cannot hold in general. Indeed $A_i(\as)$ depends on the factorisation scheme of collinear singularities, while the physical coupling should not.

We add more comments on this point, since there are conceptual analogies
(and differences) between the soft-collinear part of the Sudakov kernel and the soft limit of the DGLAP kernel. The DGLAP kernel is related to the probability of correlated emission of collinear partons with comparable values of transverse momenta
(independent collinear emission is instead taken into account through the perturbative iteration of the kernel). To obtain the DGLAP kernel, the transverse momenta are integrated up to some value of the evolution (or factorization) scale.
The transverse-momentum integral is collinear divergent in the low-momentum region:
within the \ms\ factorization scheme, the divergences are handled by using dimensional regularization in $d=4 -2 \ep$ space time dimensions, and the DGLAP kernel is defined as the coefficient of the ensuing $1/\ep$ pole
(see related comments after Eq.~(\ref{dglap})).
This is an unphysical procedure, although it is perfectly well defined for factorization purposes (a different factorization procedure would lead to a different
DGLAP kernel). In contrast, the $q_T$ integration of the Sudakov kernel does not lead to collinear divergences since the low-$q_T$ region is `physically' regularized by the definition of the measured observables. Nonetheless, the equality between the cusp anomalous dimension and the CMW coupling at ${\cal O}(\as^2)$ is not completely accidental, since at this perturbative order the \ms\ factorization procedure is equivalent to introduce a lower bound on the transverse momentum \cite{Catani:1990rr},
which practically acts as the regularization procedure that can be implemented through the use of a collinear safe observable.

In the following we introduce the all-order definitions of two soft-gluon effective couplings, and we present some perturbative results. The results are obtained by regularizing ultraviolet and infrared divergences (which are encountered at intermediate stages of the calculations) through analytic continuation in $d=4 -2\ep$ space time dimensions. Specifically, we use the customary scheme of 
conventional dimensional regularization (CDR).
The QCD bare coupling $\as^u$ and the renormalized running coupling $\as(\mu_R^2)$
in the \ms\ renormalization scheme are related by the following standard definition:
\begin{equation}
\label{asren}
\as^u \,\mu_0^{2\ep}S_\ep = \as(\mu_R^2) \,\mu_R^{2\ep} \;Z(\as(\mu_R^2);\ep)
\;\;,\quad S_\ep=(4\pi)^\ep \,e^{-\ep\gamma_E} \;\;,
\end{equation}
where $\mu_0$ is the dimensional regularization scale, $\mu_R$ is the renormalization scale and $\gamma_E$ is the Euler number. The renormalization function
$Z(\as;\ep)$ is
\begin{equation}
Z(\as,\ep)=1-\as\f{\beta_0}{\ep}+\as^2
\left(\f{\beta_0^2}{\ep^2}-\f{\beta_1}{2\ep}\right)+{\cal O}(\as^3)
\;\;,
\end{equation}
where $\beta_0$ and $\beta_1$ are the first two perturbative coefficients
of the QCD $\beta$-function $\beta(\as)$:
\beq
\label{betaf}
\beta(\as) = - \beta_0 \,\as -  \beta_1 \,\as^2 +  {\cal O}(\as^3) \;\;,
\eeq
\begin{align}
\label{betacoeff}
&12\pi \,\beta_0=11C_A-2n_F \;\;,~~~~~~~~~~~~~24\pi^2\beta_1=17C_A^2-5C_An_F-3C_Fn_F\; .
\end{align}

As we have already stated, an all-order definition of soft-gluon effective coupling has been given in Ref.~\cite{Banfi:2018mcq}. We use the same starting point as in 
Ref.~\cite{Banfi:2018mcq}. We consider a generic hard-scattering process that involves
only two massless hard partons, which can be either a $q{\bar q}$ pair 
($i=q$) or two gluons ($i=g$). We compute the probability for emitting a set of soft partons (soft gluons and soft $q{\bar q}$ pairs), and we consider the function
$w_i(k;\ep)$ that gives the `probability'\footnote{Note that this `probability' is not positive definite since it refers to the correlation part of the 
total emission probability.} of {\it correlated} emission (including the corresponding virtual corrections) of an arbitrary number of soft partons with total
momentum $k$. This function is formally defined in Eq.~(2.25) of 
Ref.~\cite{Banfi:2018mcq}, and it is 
called {\it web} function therein.

Contributions to $w_i(k;\ep)$ from virtual and real radiative corrections separately lead to ultraviolet and infrared divergences. However, the probability of correlated soft emission at fixed total momentum $k$ is a quantity that is infrared and collinear safe. Therefore, infrared singularities cancel in the computation of 
$w_i(k;\ep)$ and, after renormalization of $\as$, the soft function $w_i(k;\ep)$
is finite in the physical four-dimensional limit $\ep \to 0$. For our subsequent purposes, we consider the general $d$-dimensional function $w_i(k;\ep)$, although it is well defined at $\ep =0$.

A relevant property of $w_i(k;\ep)$ is its invariance under longitudinal boosts
along 
the direction of the momenta of the two hard partons in their centre--of--mass frame. It follows that $w_i(k;\ep)$ actually depends only on two kinematical variables: the transverse-momentum component $k_T$ of $k$ with respect to the direction of the radiating partons, and the transverse mass $m_T$ 
($m_T^2= k_T^2+k^2$). We propose the definition of two different effective couplings, 
$\ATt(\as;\ep)$ and $\Aot(\as;\ep)$, which measure the intensity of inclusive soft-parton radiation. The definitions are
\begin{align}
  \label{eq:defAT}
  \ATt(\as(\mu^2);\ep)&=\f{1}{2}\;\mu^2\int_0^\infty dm_T^2 \,dk_T^2 \;\delta(\mu^2-k_T^2) \;w_i(k;\ep) \;\;,\\
  \label{eq:defAo}
  \Aot(\as(\mu^2);\ep)&=\f{1}{2}\;\mu^2\int_0^\infty dm_T^2 \,dk_T^2 \;\delta(\mu^2-m_T^2) \;w_i(k;\ep) \;\;,
\end{align}
where $\ATt(\as;\ep=0)$ corresponds\footnote{The function $w_i(k;\ep)$ in 
Eqs.~(\ref{eq:defAT}) and (\ref{eq:defAo}) and the web function in Eq.~(2.25)
of Ref.~\cite{Banfi:2018mcq} are directly proportional, and the proportionality relation includes the overall factor $(k_T^2)^{-\ep}$ that makes $\ATt$ and $\Aot$
dimensionless in any number $d$ of dimensions.}
to the soft coupling of Ref.~\cite{Banfi:2018mcq}.

The definitions in Eqs.~(\ref{eq:defAT}) and (\ref{eq:defAo}) differ only in the
kinematical variable that is kept fixed in the integration procedure over $k$: 
$\ATt(\as(\mu^2);\ep)$ is defined at fixed value $k_T=\mu$ of the 
transverse momentum, while $\Aot(\as(\mu^2);\ep)$ is defined at fixed value
$m_T=\mu$ of the transverse mass.
In the right-hand side of Eqs.~(\ref{eq:defAT}) and (\ref{eq:defAo}), the factor
$\mu^2$ is introduced for dimensional reasons (so that $\At$ is dimensionless)
and the factor $1/2$ takes into account the fact that the integration of $w_i(k;\ep)$
includes the angular regions where the soft momentum $k$ is collinear to the momentum of each of the two hard partons. In the definitions of
Eqs.~(\ref{eq:defAT}) and (\ref{eq:defAo})
the renormalization scale $\mu_R$ is set to the value $\mu_R=\mu$.
Obviously, the soft couplings $\ATt(\as(\mu^2);\ep)$ and $\Aot(\as(\mu^2);\ep)$
are renormalization group invariant quantities, so that, at the perturbative level,
they can equivalently be expressed in terms of the running coupling $\as(\mu_R^2)$
and the ratio $\mu^2/\mu_R^2$.

The integration over $k$ in Eqs.~(\ref{eq:defAT}) and (\ref{eq:defAo}) is infrared and collinear safe, so that the limit $\ep \to 0$ is finite and well defined.
Therefore, the soft-gluon effective couplings $\AT(\as)$ and $\Ao(\as)$
in the physical four-dimensional space time are simply
\begin{equation}
\AT(\as)\equiv \ATt(\as;\ep=0) \;\;, \quad \quad \quad
\Ao(\as)\equiv \Aot(\as;\ep=0) \;\;.
\end{equation}
Nonetheless we insist in using a $d$-dimensional definition of the soft-gluon coupling
for a twofold (formal and practical) purpose. The formal aspects will be discussed below. At the practical level, the $d$-dimensional definition permits a direct application of the effective coupling in the context of hadron collisions,
where Sudakov resummation can be sensitive to the PDFs of the colliding hadrons (and the related \ms\ factorization procedure in $d$ dimensions).

The coefficients of the perturbative expansion of $\At$ and $\A$ are defined analogously to those in Eq.~(\ref{cusp}):
\beq
\label{atpert}
\At(\as;\ep) = \sum_{n=1}^\infty \left(\frac{\as}{\pi}\right)^n \At^{(n)}(\ep)\;,
\quad \quad
\A(\as) = \sum_{n=1}^\infty \left(\frac{\as}{\pi}\right)^n \A^{(n)}\;.
\eeq
The $\ep$-expansion at the $n$-th perturbative order is denoted as follows 
\beq
\label{ankcoef}
\At^{(n)}(\ep) = \A^{(n)} + \sum_{k=1}^\infty \ep^k \;\At^{(n; k)}\;.
\eeq

To make explicit the definition of the overall normalization of $\At(\as;\ep)$
(and $w_i(k;\ep)$), we report the expression of the lowest-order contribution: 
\begin{equation}
\label{at1}
    \ATt^{(1)}(\ep) = \Aot^{(1)}(\ep) = C_i \;c(\ep) \;\;,
\end{equation}
where
\beq
\label{c1}
c(\ep) \equiv \f{e^{\ep\GE}}{\Gamma(1-\ep)} = 1 - \f{\pi^2}{12}  \, \ep^2
- \f{1}{3} \,\zeta_3 \, \ep^3 + {\cal O}(\ep^4) \;\;,
\eeq
and $\Gamma(z)$ is the Euler $\Gamma$-function.
We note that the two soft couplings $\ATt^{(1)}$ and $\Aot^{(1)}$ are exactly equal at the lowest perturbative order. This equality simply follows from the fact that the 
lowest-order contribution to $w_i(k;\ep)$ is proportional to $\delta(k^2)=
\delta(m_T^2 - k_T^2)$.
We also note (see Eq.~(\ref{c1})) that the $\ep$ dependence of $\At^{(1)}(\ep)$
starts at ${\cal O}(\ep^2)$ (i.e., the coefficient $\At^{(1; 1)}$ at ${\cal O}(\ep)$
vanishes).
This mild $\ep$ dependence is of entirely `kinematical' origin
(it arises from the $d$-dimensional phase space), since (due to helicity conservation) the dynamics of soft-gluon radiation does not produce any
$\ep$ dependence at the lowest perturbative order.

We anticipate (see below) that, in the physical four-dimensional space time, both soft couplings in Eqs.~(\ref{eq:defAT}) and (\ref{eq:defAo}) are equal to the CMW coupling $\A^{CMW}$ up to ${\cal O}(\as^2)$. Therefore, we have  
\begin{equation}
\label{at2}
    \AT^{(2)} = \Ao^{(2)} = A^{(2)}_i \;\;.
\end{equation}

One of the main results of this paper is the following all-order relation between the cusp anomalous dimension $A_i(\as)$ and the soft-gluon couplings:
\begin{equation}
\label{main}
  \ATt(\as;\ep=\beta(\as)) = \Aot(\as;\ep=\beta(\as)) = A_i(\as) \;\;.
  \end{equation}
This relation can be derived in differents ways. A procedure that we have used consists in considering threshold resummation \cite{Sterman:1986aj, Catani:1989ne, Catani:1990rp}
for the production of high-mass systems in hadron collisions.
The threshold resummed cross section is related to the evolution of the PDFs in the soft limit (see Eq.~(\ref{ap})). We have applied both soft couplings in 
Eqs.~(\ref{eq:defAT}) and (\ref{eq:defAo}) to the computation of the threshold resummed cross section and we have obtained the result in Eq.~(\ref{main}).

The relation in Eq.~(\ref{main}) can be rewritten in the following form:
\beq
\label{dglap}
A_i(\as(\mu_F^2))= \frac{d}{d \ln \mu_F^2} \;{\cal P_{\ep}} 
\left\{ \int_0^{\mu_F^2} \frac{dq_T^2}{q_T^2} \;\At((\as(q_T^2);\ep) \right\}
\;\;,
\eeq
where 
$\At$ is equivalently $\ATt$ or $\Aot$, and
${\cal P_{\ep}}$ is the projection operator \cite{Curci:1980uw}
that extracts the $\ep$ poles
(in \ms\ form)
of the function of $\as(\mu_F^2)$ and $\ep$ in the curly bracket.
The equivalence between Eqs.~(\ref{main}) and (\ref{dglap}) can be proven by using some $d$-dimensional technicalities. We would like to point out that the relation between $\At$ and $A_i$ as expressed in the form of Eq.~(\ref{dglap}) is in direct correspondence with our previous qualitative discussion about the relation between the Sudakov kernel and the DGLAP kernel. The soft coupling $\At((\as(q_T^2);\ep)$
gives the intensity of the spectrum of correlated soft and collinear emission
of partons with total transverse momentum $q_T$. In the right-hand side of 
Eq.~(\ref{dglap}), the $q_T$ spectrum is integrated over the region from $q_T=0$ up
to some value of the factorization scale $\mu_F$. Following the \ms\ factorization procedure, the $\ep$ poles that arise from the $d$-dimensional regularization of the collinear singularities in the region around $q_T \simeq 0$ are then extracted to obtain (actually, to define) the intensity $A_i(\as(\mu_F^2))$ 
of soft radiation in the DGLAP kernel 
(i.e., the cusp anomalous dimension in Eq.~(\ref{ap})).

Equation (\ref{main}) relates\footnote{An equality between the soft anomalous dimension (which is related to the cusp anomalous dimension) and the $d$-dimensional rapidity anomalous dimension
at the point $\ep=\beta(\as)$ is presented in Ref.~\cite{Vladimirov:2016dll}.}
the cusp anomalous dimension to the $d$-dimensional soft-gluon coupling at the conformal point $\ep=\beta(\as)$, where the 
$d$-dimensional QCD $\beta$-function $\beta(\as) - \ep$ vanishes.
The relation (\ref{main}) is not specific of QCD, and it also applies to other gauge
theories. In particular, in the case of ${\cal N}=4$ maximally supersymmetric 
Yang--Mills theory we have $\beta(\as)=0$ and, therefore, 
the cusp anomalous dimension coincides with the physical (four-dimensional) 
soft-gluon coupling: ${{\mathscr A}}_{T}(\as)={{\mathscr A}}_{0}(\as)=A(\as)$.

According to Eq.~(\ref{main}), there is a non-trivial interplay between the perturbative dependence of the cusp anomalous dimension and the $d$-dimensional dependence of the soft-gluon coupling. In particular, since the $\ep$-dependence of 
$\At^{(1)}(\ep)$ starts at ${\cal O}(\ep^2)$ (see Eqs.~(\ref{at1}) and (\ref{c1})),
Eq.~(\ref{main}) directly implies the equivalence up to 
${\cal O}(\as^2)$ (see Eq.~(\ref{at2})) between the cusp anomalous dimension
and the four-dimensional soft-gluon coupling (or the CMW coupling).
As we have already recalled, this equivalence is not completely accidental
\cite{Catani:1990rr} and, at the purely technical level, it can be viewed as a consequence of the mild ${\cal O}(\ep^2)$ dependence in Eqs.~(\ref{at1}) and 
(\ref{c1}).

The relation in Eq.~(\ref{main}) also states that the two $d$-dimensional soft couplings, $\ATt(\as;\ep)$ and $\Aot(\as;\ep)$, become equal by setting 
$\ep=\beta(\as)$. Starting from ${\cal O}(\as^2)$
(see Eqs.~(\ref{at2ep}) and (\ref{ao2ep}) below), the $\ep$-dependence of the two soft couplings is very different. In view of this, we find it remarkable that such a different $\ep$-dependence conspires to make the coupling equal at $\ep=\beta(\as)$.
Incidentally, such a different $\ep$-dependence and the relation (\ref{main})
imply that the two four-dimensional soft couplings, $\AT(\as)$ and  $\Ao(\as)$,
inevitably differ starting from ${\cal O}(\as^3)$.
Moreover, the difference $\AT(\as) - \Ao(\as)$ is necessarily due to perturbative contributions that are proportional to the coefficients, $\beta_0, \beta_1$ and so forth, of the QCD $\beta$-function.

In addition to be interesting for its intrinsic structure,
the relation in Eq.~(\ref{main}) can be exploited for several different purposes.
It can be used to crosscheck explicit perturbative computations of $A_i(\as)$ and
$\At(\as;\ep)$. Once one the the three functions $A_i, \ATt$ and $\Aot$ is known
at some perturbative order, Eq.~(\ref{main}) can exploited to extract information on the other two functions (in the following we explicitly make this use of 
Eq.~(\ref{main})). The relation (\ref{main}) can also be used to obtain the cusp anomalous dimension $A_i(\as)$ through the $d$-dimensional perturbative calculation of one of the two soft couplings $\At(\as;\ep)$.

We have computed the soft function $w_i(k;\ep)$ at ${\cal O}(\as^2)$ by combining the one-loop correction to single soft-gluon radiation \cite{Catani:2000pi}
with the $d$-dimensional integration of double soft-parton radiation at the tree level
\cite{Catani:1999ss}.
Then, using Eqs.~(\ref{eq:defAT}) and (\ref{eq:defAo}), we have computed the 
soft-gluon effective couplings in $d$ dimensions at ${\cal O}(\as^2)$, and we obtain the following results \cite{inprep}.
In the case of $\ATt(\as;\ep)$ we find
\beeq
\label{at2ep}
 \ATt^{(2)}(\ep)&=& C_i \;\Bigl\{  
    - \frac{ c(\ep)\, (11 C_A-2 n_F)}{12\,
    \ep} 
    +  \frac{ c(2\ep)\, \pi }{\sin (\pi  \ep) }
    \frac{ [ C_A (11- 7 \ep)-2\,n_F (1-\ep)]}{4 (3-2 \ep)(1-2 \ep)} 
   \Bigr.
   \nn \\
   &+& \Bigl.
    \frac{ C_A\, c(2\ep) \, h(\ep)\, \pi }{2 \sin (\pi  \ep)
    }- \frac{ C_A\,  c(2\ep)\, \pi^2 }{2 \sin^2 (\pi  \ep)} 
   \left( \frac{2- \sin^2 (\pi  \ep)}{\cos(\pi  \ep) } 
   - \frac{2 \sin (\pi  \ep)}{\pi  \ep}\right)
   \Bigr\} \;\;,
\eeeq
where
\beq
h(\ep) =
\gamma_E + \psi(1 - \ep) + 2\, \psi(1 + 2 \ep) - 2\, \psi(1 + \ep) \;\;,
\eeq
and $\psi(1+z)=\frac{d \ln \Gamma(1+z)}{dz}$.
In the case of $\ATt(\as;\ep)$ we find
\beeq
\label{ao2ep}
 \Aot^{(2)}(\ep)&=& C_i \;\Bigl\{  
    - \frac{ c(\ep)\, (11 C_A-2 n_F)}{12\,
    \ep} 
    +  \frac{ c^2(2\ep) }{\ep \, c^2(\ep)}
    \frac{ [ C_A (11- 7 \ep)-2\,n_F (1-\ep)]}{4 (3-2 \ep)(1-2 \ep)} 
   \Bigr.
   \nn \\
   &+& \Bigl.
   \frac{ C_A \,c^2(2\ep) \,r(\ep)}{2 (1-2\ep) \,c^2(\ep)} 
   - \frac{ C_A\,  c(2\ep)}{2 \,\ep^2} 
   \left( \frac{(\pi \ep)^2 \cos(\pi  \ep)}{\sin^2 (\pi  \ep)}
   +\frac{\pi \,\ep}{\sin(\pi  \ep)} - \frac{2 \,c(2\ep)}{c^2(\ep)}
   \right)
   \Bigr\} \;\;,
\eeeq
where
\beq
r(\ep) = \frac{2}{1+\ep}\,\,  _3F_2(1,1,1-\epsilon;2-2\epsilon,2+\epsilon;1)
       - \frac{1}{1-\ep}\,\,  _3F_2(1,1,1-\epsilon;2-2\epsilon,2-\epsilon;1) \;\;,
\eeq
and $_3F_2(\alpha, \beta, \gamma; \delta, \rho;z)$ is the generalized hypergeometric function of the variable $z$.

The $\ep$-expansion up to ${\cal O}(\ep^2)$ of the 
second-order expressions in Eqs.~(\ref{at2ep}) and (\ref{ao2ep}) gives
\beeq
\label{at2ep2}
 \ATt^{(2)}(\ep)&=&A_i^{(2)} +\ep\, C_i \left[C_A \left( \frac{101}{27} - \frac{11\, \pi^2}{144} -\frac{7 \zeta_3}{2}\right) 
+  n_F \left( \frac{\pi^2}{72}  -\frac{14}{27}  \right) \right] 
  \\
&+&\ep^2\, C_i \left[ C_A \left( \frac{607}{81} - \frac{67\, \pi^2}{216} -\frac{77 \zeta_3}{36} - \frac{7\, \pi^4}{120}\right) 
+  n_F \left( \frac{5\,\pi^2}{108}  -\frac{82}{81} + \frac{7 \zeta_3}{18} \right) \right] +{\cal{O}}(\ep^3) \;\;,\nn
\eeeq
\beeq
\label{ao2ep2}
   \Aot^{(2)}(\ep)&=&A_i^{(2)} +\ep\, C_i \left[C_A \left( \frac{101}{27} - \frac{55\, \pi^2}{144} -\frac{7 \zeta_3}{2}\right) 
+  n_F \left( \frac{5\, \pi^2}{72}  -\frac{14}{27}  \right) \right]
  \\
&+&\ep^2\, C_i \left[C_A \left( \frac{607}{81} - \frac{67\, \pi^2}{72} -\frac{143 \zeta_3}{36} - \frac{\pi^4}{36}\right) 
+  n_F \left( \frac{5\,\pi^2}{36}  -\frac{82}{81} + \frac{13 \zeta_3}{18} \right) \right] +{\cal{O}}(\ep^3) \;\;,\nn
\eeeq
where $A_i^{(2)}$ is given in Eq.~(\ref{a2cusp}).
From these equations we see that the $\ep$ dependence of the two soft couplings
$\ATt^{(2)}(\ep)$ and $\Aot^{(2)}(\ep)$ is already different at ${\cal O}(\ep)$.
We also see that the limit $\ep \to 0$ of our explicit calculation at 
${\cal O}(\as^2)$ leads to the equality in Eq.~(\ref{at2}) between the two soft couplings and the CMW coupling.
At the computational level the equality  $\AT^{(2)} = \Ao^{(2)}$ originates as follows. Since $\ATt^{(1)}(\ep) = \Aot^{(1)}(\ep)$, the value of $\At^{(2)}(\ep)$
at $\ep=0$ is determined by the behaviour of the soft function $w_i(k;\ep)$ 
in the region where $k^2 \simeq 0$. In this region we have $m_T^2 \simeq k_T^2$ and, therefore, 
the difference between the right-hand side of 
Eqs.~(\ref{eq:defAT}) and (\ref{eq:defAo}) (and, hence, between
the two soft couplings) is not effective.

We now present our computation of the third-order coefficients 
$\AT^{(3)}$ and $\Ao^{(3)}$ of both four-dimensional soft couplings.
To this purpose we use Eq.~(\ref{betaf}) and we perturbatively expand Eq.~(\ref{main})
in terms of the coefficients $\At^{(n; k)}$ that are defined in Eq.~(\ref{ankcoef}).
We obtain
\beq
\label{a3vscusp}
  \A^{(3)} = A_i^{(3)} -  (\beta_0 \pi)^2 \;\At^{(1; 2)} + (\beta_0 \pi) \;\At^{(2; 1)} \;\;.
\eeq
This relation applies to both soft couplings $\ATt$ and $\Aot$
(we have omitted the corresponding subscripts $T$ and $0$), and we have also used  
$\At^{(1; 1)} = 0$ (see Eqs.~(\ref{at1}) and (\ref{c1})).
Since we have determined $\At^{(1)}(\ep)$ and $\At^{(2)}(\ep)$ to all orders in the 
$\ep$-expansion, the explicit values of the coefficients $\At^{(1; 2)}$
and $\At^{(2; 1)}$ can be directly read from Eqs.~(\ref{at1}), (\ref{c1}), 
(\ref{at2ep2}) and (\ref{ao2ep2}). Inserting these coefficients in 
Eq.~(\ref{a3vscusp}) we can explicitly relate $\A^{(3)}$ to the coefficient 
$A_i^{(3)}$ (see Eq.~(\ref{a3cusp})) of the cusp anomalous dimension. We obtain the following results:
\begin{align}
\label{a3t}
  \AT^{(3)}&=A_i^{(3)}+ C_i \,(\beta_0 \pi)^2 \;\frac{\pi^2}{12} 
+ C_i  \,(\beta_0 \pi) 
\left[ C_A \left( \frac{101}{27} - \frac{11\, \pi^2}{144} -\frac{7 \zeta_3}{2}\right) 
+  n_F \left( \frac{\pi^2}{72}  -\frac{14}{27}  \right) \right] \;\;,
\end{align}
\begin{align}
\label{a3o}
  \Ao^{(3)}&=A_i^{(3)}+ C_i \,(\beta_0 \pi)^2 \;\frac{\pi^2}{12} 
+ C_i  \,(\beta_0 \pi) 
\left[ C_A \left( \frac{101}{27} - \frac{55\, \pi^2}{144} -\frac{7 \zeta_3}{2}\right) 
+  n_F \left( \frac{5\, \pi^2}{72}  -\frac{14}{27}  \right) \right] \;\;.
\end{align}
Our result in Eq.~(\ref{a3t}) for the third-order coefficient of the soft-gluon coupling $\AT(\as)$ agrees with the corresponding result presented in 
Ref.~\cite{Banfi:2018mcq} (see Eqs.~(3.9) and (3.11b) therein).

Using the value of $A_i^{(3)}$ in Eq.~(\ref{a3cusp}), the results in Eqs.~(\ref{a3t}) and (\ref{a3o})
explicitly relate the four-dimensional (physical) soft-gluon effective couplings 
$\AT(\as)$ and $\Ao(\as)$ with the \ms\ renormalized coupling $\as$ up to
${\cal O}(\as^3)$. This relation generalizes the ${\cal O}(\as^2)$ CMW relation
in Eq.~(\ref{cmw})
to the third order, and it can be used to construct the Sudakov kernel for soft-gluon resummation of infrared and collinear safe observables at NNLL accuracy 
\cite{Banfi:2018mcq}.

In the case of the soft-gluon coupling $\Ao(\as)$ we have also computed its relation with the \ms\ coupling at ${\cal O}(\as^4)$. More precisely, we obtain an explicit relation between $\Ao^{(4)}$ and the corresponding coefficient 
$A_i^{(4)}$ 
of the cusp anomalous dimension.
We find
  \begin{align}
\label{a4o}
\Ao^{(4)}&=A_i^{(4)} + C_i \biggl\{  C_A^3 \left(\frac{121 \pi ^2 \zeta_3}{288}-\frac{21755
    \zeta_3}{864}+\frac{33 \zeta_5}{4}+\frac{847 \pi ^4}{17280}-\frac{41525
    \pi ^2}{15552}+\frac{3761815}{186624}\right) \nn \\
    &+C_A^2 n_F 
    \left(-\frac{11 \pi ^2 \zeta_3}{144}+\frac{6407 \zeta_3}{864}-\frac{3
    \zeta_5}{2}-\frac{11 \pi ^4}{432}+\frac{9605 \pi
    ^2}{7776}-\frac{15593}{1944}\right) \nn \\
    &+C_A C_F n_F 
    \left(\frac{17 \zeta_3}{9}+\frac{11 \pi ^4}{1440}+\frac{55 \pi
    ^2}{576}-\frac{7351}{2304}\right)+ C_A n_F^2 \left(-\frac{179
    \zeta_3}{432}+\frac{13 \pi ^4}{4320}-\frac{695 \pi
    ^2}{3888}+\frac{13819}{15552}\right)\nn \\
     &+C_F n_F^2 \left(-\frac{19
    \zeta_3}{72}-\frac{\pi ^4}{720}-\frac{5 \pi
    ^2}{288}+\frac{215}{384}\right)+n_F^3
    \left(-\frac{\zeta_3}{108}+\frac{5 \pi ^2}{648}-\frac{29}{1458}\right) \biggl\}
   \;\;.
\end{align}
This fourth-order result can be used for applications to
soft-gluon resummed calculations of infrared and collinear safe observables at
the next-to-next-to-next-to-leading logarithmic (N$^3$LL) accuracy. 

Knowing the result in Eq.~(\ref{a4o}) and exploiting the relation in Eq.~(\ref{main}),
we can also explicitly determine the third-order coefficient $\Aot^{(3)}(\ep)$
of the $d$-dimensional soft coupling at ${\cal O}(\ep)$. To illustrate the procedure,
we perturbatively expand Eq.~(\ref{a4o}) in terms of the coefficients
$\At^{(n; k)}$ of Eq.~(\ref{ankcoef}), and we obtain
\beq
\label{a4vscusp}
  \Ao^{(4)} = A_i^{(4)} + (\beta_0 \pi) \;\Aot^{(3; 1)} -  (\beta_0 \pi)^2 \;\Aot^{(2; 2)}
+ (\beta_1 \pi^2) \;\Aot^{(2; 1)} + (\beta_0 \pi)^3 \;\Aot^{(1; 3)}
- 2 (\beta_1 \beta_0 \pi^3) \;\Aot^{(1; 2)}
\;\;,
\eeq
where we have used $\Aot^{(1; 1)}=0$. The explicit coefficients 
$\Aot^{(1; 2)}$ and $\Aot^{(1; 3)}$ at the first order and
$\Aot^{(2; 1)}$ and $\Aot^{(2; 2)}$ at the second order
can be read from Eqs.~(\ref{at1}), (\ref{c1}) and Eq.~(\ref{ao2ep2}), respectively.
Therefore, by comparing Eqs.~(\ref{a4o}) and (\ref{a4vscusp}) we obtain
\beq
  \Aot^{(3)}(\ep) = \Ao^{(3)} + \ep \;\Aot^{(3; 1)} + {\cal O}(\ep^2) \;\;,
\eeq
with the explicit result
\begin{align}
   \Aot^{(3;1)} &= C_i \biggl\{  C_A^2 \left(\frac{11 \pi ^2 \zeta_3}{24}-\frac{225 \zeta_3}{8}+9
    \zeta_5+\frac{121 \pi ^4}{4320}-\frac{4651 \pi
    ^2}{1296}+\frac{403861}{15552}\right) \nn \\
    &+C_A n_F  \left(\frac{289
    \zeta_3}{72}-\frac{29 \pi ^4}{2160}+\frac{2717 \pi
    ^2}{2592}-\frac{48241}{7776}\right)+C_F n_F  \left(\frac{19
    \zeta_3}{12}+\frac{\pi ^4}{120}+\frac{7 \pi
    ^2}{96}-\frac{1711}{576}\right) \nn \\
    &+n_F^2
    \left(-\frac{\zeta_3}{18}-\frac{5 \pi ^2}{72}+\frac{70}{243}\right) \biggl\} 
    \;\;.
  \end{align}

We comment on our derivation of the result in Eq.~(\ref{a4o}).
The soft-gluon effective coupling $\Aot(\as;\ep)$ is particularly suitable
in the context of threshold resummation \cite{Sterman:1986aj, Catani:1989ne, Catani:1990rp}
for the production of colourless high-mass systems in hadron collisions.
The threshold resummed cross section for these processes is presently known
in explicit form up to N$^3$LL accuracy \cite{Vogt:2000ci}-\cite{Bonvini:2014tea}.
We have applied $\Aot(\as;\ep)$ to threshold resummation and, exploiting the known 
N$^3$LL results \cite{Catani:2014uta}, we obtain Eq.~(\ref{a4o}).

The result in Eq.~(\ref{a4o}) relates the fourth-order perturbative term $\Ao^{(4)}$
of the soft coupling $\Ao(\as)$ to the corresponding term $A_i^{(4)}$ of the cusp anomalous dimension $A_i(\as)$. Since $A_i^{(4)}$ is not fully known in analytic form,
we add some comments on the fourth-order results.

We have examined the colour structure of soft multiparton radiation from two hard partons at ${\cal O}(\as^4)$ and, consequently, we can obtain the general colour structure of the soft function $w_i(k;\ep)$ or, equivalently 
(due to Eqs.~(\ref{eq:defAT}) and (\ref{eq:defAo})),
the colour structure 
of the soft coupling.
We write this structure in the following form:
\beq
\label{a4i}
\At^{(4)}(\ep) = C_i \;\Anoti^{(4)}_{[2]}(\ep) 
+ \frac{d_{Ai}^{(4)}}{N_i} \;\Anoti^{(4)}_{[4A]}(\ep)
+ n_F \frac{d_{Fi}^{(4)}}{N_i} \;\Anoti^{(4)}_{[4F]}(\ep) \;\;,
\eeq
where $N_i$ is the dimension of the colour representation of the hard parton $i$
($N_i=N_A= N_c^2-1$ if $i=g$, and  $N_i=N_F= N_c$ if $i=q,{\bar q}$), and
$d_{xy}^{(4)}$ are the quartic Casimir invariants (we use the normalization of
$d_{xy}^{(4)}$ as in Eqs.~(2.6)--(2.10) of Ref.~\cite{Moch:2018wjh}).
The entire dependence of $\At^{(4)}(\ep)$ (for both couplings
$\ATt^{(4)}(\ep)$ and $\Aot^{(4)}(\ep)$) on the colour of the hard parton $i$ is embodied in the Casimir dependent factors that we have explicitly written in the right-hand side of Eq.~(\ref{a4i}). The `quartic' ($\Anoti^{(4)}_{[4A]}(\ep)$ and
$\;\Anoti^{(4)}_{[4F]}$) and `quadratic' ($\Anoti^{(4)}_{[2]}$)
coefficients do not depend on the type of radiating parton~$i$. In particular,
$\Anoti^{(4)}_{[4A]}(\ep)$ and $\;\Anoti^{(4)}_{[4F]}$ are colour blind (they do not depend on $N_c$ and $n_F$). The coefficient $\Anoti^{(4)}_{[2]}$ still depends on 
$N_c$ and $n_F$, and this dependence involves all the colour structures that appear in the curly bracket of Eq.~(\ref{a4o})
plus an additional term with colour factor $C_F^2 n_F$.

The presence in Eq.~(\ref{a4i}) of the quartic Casimir invariants violates Casimir scaling (i.e., the proportionality relation $\At \propto C_i$). Nonetheless 
$\At^{(4)}$ in Eq.~(\ref{a4i}) still fulfils a form of {\it generalized} Casimir scaling (in terms of three colour coefficients that depend on $i$) since 
$\Anoti^{(4)}_{[2]}, \Anoti^{(4)}_{[4A]}$ and $\Anoti^{(4)}_{[4F]}$ do not depend on the hard parton $i$.

Setting $\ep=0$ in Eq.~(\ref{a4i}) and using Eq.~(\ref{main}), we obtain the colour structure of the four-dimensional soft coupling $\Ao^{(4)}$ 
(or, analogously\footnote{The expression in Eq.~(\ref{4sc}) is equally valid for the soft coupling $\AT^{(4)}$ through the replacement 
$\Anott^{(4)}_{0 [2]} \to \Anott^{(4)}_{T [2]}$.}, 
$\AT^{(4)}$) and of the cusp anomalous dimension $A_i^{(4)}$:
\beq
\label{4sc}
\Ao^{(4)} = C_i \;\Anott^{(4)}_{0 [2]} 
+ \frac{d_{Ai}^{(4)}}{N_i} \;A^{(4)}_{[4A]}
+ n_F \frac{d_{Fi}^{(4)}}{N_i} \;A^{(4)}_{[4F]} \;\;,
\eeq
\beq
\label{4cc}
A_i^{(4)} = C_i \;A^{(4)}_{[2]} 
+ \frac{d_{Ai}^{(4)}}{N_i} \;A^{(4)}_{[4A]}
+ n_F \frac{d_{Fi}^{(4)}}{N_i} \;A^{(4)}_{[4F]} \;\;,
\eeq
where, analogously to Eq.~(\ref{a4i}), the full dependence on the colour of the hard parton $i$ is entirely controlled by the Casimir dependent coefficients
$C_i, d_{Ai}^{(4)}/N_i$ and $d_{Fi}^{(4)}/N_i$.

We note that, to obtain Eqs.~(\ref{4sc}) and (\ref{4cc}) from Eq.~(\ref{a4i}), we have exploited Eq.~(\ref{main}) and the property that the difference 
$\Ao^{(4)} - A_i^{(4)}$ fulfils Casimir scaling (see Eq.~(\ref{a4vscusp})), since
the perturbative terms
$\Aot^{(n)}(\ep)$ with $n=1,2,3$ fulfil Casimir scaling. In particular, in 
Eqs.~(\ref{4sc}) and (\ref{4cc}) we have set $\Anoti^{(4)}_{ 0 [2]}(\ep=0) \equiv
\Anott^{(4)}_{0 [2]}$, then we have related the `quadratic' coefficients of the soft coupling ($\Anott^{(4)}_{0 [2]})$) and of the cusp anomalous dimension 
($A^{(4)}_{[2]}$) through Casimir scaling:
\beq
\label{4scaling}
\Ao^{(4)} - A_i^{(4)} = C_i \left( \Anott^{(4)}_{0 [2]} -  A^{(4)}_{[2]} \right) \;\;,
\eeq
and, finally, we have derived and implemented the following equalities
\beq
\Anoti^{(4)}_{ 0 [4A]}(\ep=0) \equiv \Anott^{(4)}_{0 [4A]} = A^{(4)}_{[4A]} \;\;,
\quad 
\Anoti^{(4)}_{ 0 [4F]}(\ep=0) \equiv \Anott^{(4)}_{0 [4F]} = A^{(4)}_{[4F]} \;\;,
\eeq
between the `quartic' coefficients of the soft coupling ($\Anott^{(4)}_{0 [4A]},
\Anott^{(4)}_{0 [4F]}$) and of the cusp anomalous dimension
($A^{(4)}_{[4A]},  A^{(4)}_{[4F]}$).
Our result in Eq.~(\ref{a4o}) is fully consistent with the Casimir scaling relation in Eq.~(\ref{4scaling}).

We note that the generalized Casimir scaling of the soft coupling in Eq.~(\ref{a4i})
and the relation in Eq.~(\ref{main}) necessarily imply
the same scaling for the cusp anomalous dimension in Eq.~(\ref{4cc}).
The generalized Casimir scaling of the cusp anomalous dimension has been 
conjectured and verified to good numerical accuracy in 
Ref.~\cite{Moch:2018wjh}.
We also note that at the fourth order the DGLAP kernel $P_{gg}(\as;z)$ includes a contribution with the quartic Casimir invariant $d_{FF}^{(4)}$, which is absent in
$A_g^{(4)}$ of Eq.~(\ref{4cc}). Such contribution to $P_{gg}^{(4)}(\as;z)$ vanishes in the soft limit, consistently with the approximate numerical result of 
Ref.~\cite{Moch:2018wjh}.

The fourth-order term $A_i^{(4)}$ of the cusp anomalous dimension is not yet known in full analytic form, although it is known with good numerical accuracy.
The analytic results, which regard the coefficients of various colour factors, have been obtained by using different methods: the computation of the soft limit of the DGLAP kernel \cite{Gracey:1994nn, Davies:2016jie, Moch:2017uml},
the fourth-order evaluation of form factors \cite{Henn:2016men, Lee:2019zop, Henn:2019rmi},
the cusp renormalization of Wilson loops \cite{Beneke:1995pq}
(as we have previously observed, the relation (\ref{main}) leads to another method
to compute $A_i(\as)$ through the evaluation of the $d$-dimensional soft coupling
$\At(\as;\ep)$). In particular, the `quartic' coefficient  $A^{(4)}_{[4F]}$
in Eqs.~(\ref{4sc}) and (\ref{4cc}) has been computed very recently 
\cite{Lee:2019zop, Henn:2019rmi}.
The coefficients of the remaining colour factor contributions to $A_i^{(4)}$
have been evaluated in approximate numerical form \cite{Moch:2018wjh}.

The quantitative effect on the soft coupling $\Ao^{(4)}$ of the present numerical uncertainty of $A_i^{(4)}$ is {\it very} small, since the quantitative value 
of $\Ao^{(4)}$ turns out to be dominated by the contribution
$\Ao^{(4)} - A_i^{(4)}$ that we have explicitly computed in Eq.~(\ref{a4o}).
To see this, we write
\beq
\Ao^{(4)} = \left( \Ao^{(4)} - A_i^{(4)} \right) \left[ \,1 + \Delta_i \,\right]
\;\;, \quad \quad \Delta_i \equiv \frac{A_i^{(4)}}{\Ao^{(4)} - A_i^{(4)}} \;\;.
\eeq
The term $\Delta_i$ depends on $n_F$. Using $\Ao^{(4)} - A_i^{(4)}$ from 
Eq.~(\ref{a4o}) and $A_i^{(4)}$ from Ref.~\cite{Moch:2018wjh}
and setting $n_F=5$ (with $N_c=3$) we obtain
\beq
\Delta_i(n_F=5) = \bigl( -0.222(5) \,\delta_{i q} + 4.05(4) \,\delta_{i g}\bigr) 
\times 10^{-2} \;\;,
\eeq
where the numbers in brackets indicate the numerical uncertainty (due to
$A_i^{(4)}$ \cite{Moch:2018wjh}) of the preceding digit. Similar quantitative results
are obtained for $n_F=3,4$. The term $\Delta_i$ turns out to contribute to 
$\Ao^{(4)}$ at the level of few percents, so that a small uncertainty on 
$A_i^{(4)}$ leads to a very small uncertainty on $\Ao^{(4)}$.

As observed in Ref.~\cite{Moch:2018wjh}, due to the actual values of the `quartic' coefficients $A^{(4)}_{[4A]}$ and $A^{(4)}_{[4F]}$ in Eq.~(\ref{4cc}), numerical Casimir scaling is completely broken in the fourth-order term $A_i^{(4)}$
of the cusp anomalous dimension. However, due to the smallness of $\Delta_i$,
the soft coupling $\Ao^{(4)}$ still fulfils numerical Casimir scaling
($\Ao^{(4)} \propto C_i $) modulo corrections at the few percent level.

We report the numerical value of the soft coupling $\Ao(\as)$ with $N_c=3$
up to ${\cal O}(\as^4)$. Using $\Ao^{(3)}$ from Eq.~(\ref{a3o}),
$\Ao^{(4)} - A_i^{(4)}$ from Eq.~(\ref{a4o}), $A_i^{(4)}$ (with its numerical uncertainty) from Ref.~\cite{Moch:2018wjh} and setting $n_F=5$, we have
\beq
\label{a4num}
\Ao(\as) = C_i \frac{\as}{\pi} \Bigl[ 1 + 0.54973\, \as - 1.7157 \, \as^2
- \bigl( 5.9803(3) \,\delta_{i q} + 6.236(2) \,\delta_{i g}
\bigr) \as^3 + {\cal O}(\as^4)
\Bigr] \;\;.
\eeq
The perturbative expansion in Eq.~(\ref{a4num}) can be compared with the corresponding
perturbative expansion of the cusp anomalous dimension in Eq.~(4.4) of 
Ref.~\cite{Moch:2018wjh}. From the comparison we can see that the 
third-order\footnote{For comparison with the value $1.7157$ in Eq.~(\ref{a4num}),
we note that the numerical value of the
third-order coefficient (see Eq.~(\ref{a3t})) of the soft coupling $\AT(\as)$
is 0.49121 .}
and fourth-order numerical coefficients in $\Ao(\as)$ are sizeably larger than those
in $A_i(\as)$. Nonetheless the perturbative expansion of $\Ao(\as)$ is still numerically well behaved. We also see that the violation of Casimir scaling in the fourth-order term of $\Ao(\as)$ is numerically at the 4\% level.

We add some general (though brief) comments on the soft-gluon 
effective coupling and Sudakov resummation.

The resummation procedure of logarithmic contributions of Sudakov type requires proper kinematical approximations of the phase space for multiparton final-state radiation. Such approximations are specific of the physical observables under consideration. As a consequence, the use of one or the other of the two soft-gluon couplings $\ATt$ and $\Aot$ can be more appropriate depending on the observables.
The two soft couplings can alternatively (or equivalently) be used for the resummation treatment of different classes of observables. Some observables can also require a combined use of both soft couplings. In Ref.~\cite{Banfi:2018mcq}
the soft-gluon coupling $\AT$ has been explicitly applied to the resummation of a wide class of observables. As we have previously mentioned, the 
soft-gluon coupling $\Aot$ is particularly suitable in the context of threshold resummation and related observables, and its application to other classes of observables can be investigated.

The soft-gluon coupling $\At$ controls the intensity of the spectrum of soft and collinear radiation in the Sudakov kernel. The Sudakov kernel has other dynamical components that, roughly speaking, are due to soft non-collinear (i.e., wide-angle)
radiation and hard (i.e., non-soft) collinear radiation. Both components have to be included in a resummed calculation (see, e.g., Refs.~\cite{Banfi:2003je, Bonciani:2003nt, Banfi:2004yd} at NLL accuracy
and Refs.~\cite{Banfi:2014sua, Banfi:2018mcq} at NNLL accuracy), and their inclusion has
to be properly performed (i.e., properly matched) according to the soft coupling 
(either $\ATt$ or $\Aot$) that is specifically used in the soft-collinear component.
However, we note that, at a given fixed perturbative order (say, $\as^n$) in the Sudakov kernel, the soft-collinear component is logarithmically enhanced
(by at least one power of log) with respect to the two other components.
Therefore, the Sudakov kernel at N$^k$LL accuracy requires the knowledge of the soft coupling $\A(\as)$ up to ${\cal O}(\as^{k+1})$ and the computation of the other components up to ${\cal O}(\as^{k})$ (i.e., one order lower than the soft coupling).
For instance, to achieve NNLL accuracy in the Sudakov kernel, the third-order results
in Eqs.~(\ref{a3t}) and (\ref{a3o})
for the soft coupling have to be combined with the calculation at ${\cal O}(\as^{2})$
of the other dynamical components.

A final comment regards the process dependence of the Sudakov kernel. The soft couplings in Eqs.~(\ref{eq:defAT}) and (\ref{eq:defAo})
are computed by considering soft-parton radiation from two hard partons in a colour singlet configuration. Soft-gluon radiation 
in processes that involve several hard partons is definitely more complex than in the case of two hard-parton processes. This complex structure of soft-gluon radiation has to be taken properly into account. However, this does not affect the soft coupling 
$\At$, since $\At$ measures the intensity of radiation that is both soft and collinear to parton $i$. The complex structure of soft radiation in multiparton hard scattering only affects the soft wide-angle component of the Sudakov kernel
(see, e.g., Refs.~\cite{Banfi:2003je, Bonciani:2003nt, Banfi:2004yd} at NLL accuracy).
 
We conclude the paper with a brief summary of its content. 
We have considered the all-order extension of the CMW effective coupling in the context of soft-gluon resummation beyond NLL accuracy. We have argued that there is no unique all-order extension, namely, no extension that shares all the universality (i.e., observable-independent) features of the CMW coupling at ${\cal O}(\as^2)$. Starting from the emission probability of an arbitrary number of soft partons, we have introduced the definition in $d=4-2\ep$ space-time dimensions of two effective couplings,
$\ATt(\as;\ep)$ and $\Aot(\as;\ep)$, which measure the intensity of the inclusive spectrum for soft and collinear radiation from a massless hard parton 
$i$ ($i=q,{\bar q},g$). We have found that, to all perturbative orders, the two soft couplings are equal 
if they are evaluated
at the $d$-dimensional point 
$\ep=\beta(\as)$, and they coincide with the (four-dimensional) cusp anomalous dimension $A_i(\as)$. The limit $\ep \to 0$ is smooth and it can be used to define the four-dimensional (`physical') couplings $\AT(\as)$ and $\Ao(\as)$. The coupling $\AT(\as)$ has originally been defined in Ref.~\cite{Banfi:2018mcq}, and its explicit relation with $\as$ up to 
${\cal O}(\as^3)$ has been presented therein. We have computed both couplings,
$\AT(\as)$ and $\Ao(\as)$, up to ${\cal O}(\as^3)$ and, in the case of $\AT(\as)$,
our independent calculation confirm the result in Ref.~\cite{Banfi:2018mcq}.
In the case of $\Aot(\as;\ep)$ we are able to compute its third-order contribution up to ${\cal O}(\ep)$ and, in the four-dimensional limit,  we obtain an explicit relation
at ${\cal O}(\as^4)$ between $\Ao(\as)$ and the cusp anomalous dimension $A_i(\as)$. Moreover, we have presented
the explicit $d$-dimensional results (e.g., to all orders in the $\ep$ expansion)
for both soft couplings up to ${\cal O}(\as^2)$.

{\bf Acknowledgements}. We would like to thank Bryan Webber for comments on the manuscript. This work is supported in part by the Swiss National Science Foundation (SNF) under contracts 200020\_169041 and IZSAZ2\_173357, by MINCyT under contract SUIZ/17/05, by Conicet and by ANPCyT.

\end{document}